\documentclass[twocolumn,pre,aps,showpacs,superscriptaddress,floatfix]{revtex4-1}
\usepackage[latin1]{inputenc}
\usepackage{bm}
\usepackage[usenames]{color}
\usepackage{multirow}
\usepackage{amssymb}
\usepackage{amsbsy}
\usepackage{amsmath}
\usepackage{stmaryrd}
\usepackage{graphicx}
\usepackage{epsfig}
\usepackage{placeins}
\usepackage[normalem]{ulem}
\usepackage{bbold}
\usepackage{braket}
\usepackage{blindtext}
\usepackage{filecontents}
\usepackage{placeins}

\newcommand {\thubbard}{t_{\mathrm{h}}}
\newcommand {\figref}[1]{Fig.~\protect\ref{#1}}
\usepackage{scalerel}
\usepackage{tikz}
\usetikzlibrary{calc}
\usetikzlibrary{patterns}
\usetikzlibrary{svg.path}
\definecolor{orcidlogocol}{HTML}{A6CE39}
\tikzset{
  orcidlogo/.pic={
    \fill[orcidlogocol] svg{M256,128c0,70.7-57.3,128-128,128C57.3,256,0,198.7,0,128C0,57.3,57.3,0,128,0C198.7,0,256,57.3,256,128z};
    \fill[white] svg{M86.3,186.2H70.9V79.1h15.4v48.4V186.2z}
                 svg{M108.9,79.1h41.6c39.6,0,57,28.3,57,53.6c0,27.5-21.5,53.6-56.8,53.6h-41.8V79.1z M124.3,172.4h24.5c34.9,0,42.9-26.5,42.9-39.7c0-21.5-13.7-39.7-43.7-39.7h-23.7V172.4z}
                 svg{M88.7,56.8c0,5.5-4.5,10.1-10.1,10.1c-5.6,0-10.1-4.6-10.1-10.1c0-5.6,4.5-10.1,10.1-10.1C84.2,46.7,88.7,51.3,88.7,56.8z};
  }
}

\newcommand\orcid[1]{\href{https://orcid.org/#1}{\mbox{\scalerel*{
\begin{tikzpicture}[yscale=-1,transform shape]
\pic{orcidlogo};
\end{tikzpicture}
}{|}}}}

\usepackage[colorlinks,linkcolor=blue,citecolor=blue,urlcolor=blue]{hyperref}

\makeatletter

\begin{document}

\title{Nontrivial damping of quantum many-body dynamics}

  \author{Tjark Heitmann \orcid{0000-0001-7728-0133}}
  \email{tjark.heitmann@uos.de}
  \affiliation{Department of Physics, University of Osnabr\"uck, D-49069
    Osnabr\"uck, Germany}
  
  \author{Jonas Richter \orcid{0000-0003-2184-5275}}
  \email{j.richter@ucl.ac.uk}
  \affiliation{Department of Physics and Astronomy, University College London,
    Gower Street, London WC1E 6BT, UK}
  
  \author{Jochen Gemmer}
  \email{jgemmer@uos.de}
  \affiliation{Department of Physics, University of Osnabr\"uck, D-49069
    Osnabr\"uck,
    Germany}
  
  \author{Robin Steinigeweg \orcid{0000-0003-0608-0884}}
  \email{rsteinig@uos.de}
  \affiliation{Department of Physics, University of Osnabr\"uck, D-49069
    Osnabr\"uck, Germany}
  
  \date{\today}
  
  
  \begin{abstract}
    Understanding how the dynamics of a given quantum system with 
    many degrees of freedom is altered by the presence of a generic perturbation 
    is a notoriously difficult question. Recent works predict that, in the 
    overwhelming majority of cases, the unperturbed dynamics is just damped by a 
    simple function, e.g., exponentially as expected from Fermi's golden rule. 
    While these predictions rely on random-matrix arguments and typicality, they 
    can only be verified for a specific physical situation by comparing to the 
    actual solution or measurement. Crucially, it also remains unclear how frequent 
    and under which conditions counterexamples to the typical behavior occur. In 
    this work, we discuss this question from the perspective of projection-operator 
    techniques, where exponential damping of a density
    matrix occurs in the interaction picture but not {\it necessarily} in the 
    Schr\"odinger picture. We show that a nontrivial damping in 
    the 
    Schr\"odinger picture can emerge if the dynamics in the unperturbed system 
    possesses rich features, for instance due to the presence of
    strong interactions. This suggestion has consequences for
    the time dependence of correlation functions.
    We substantiate our theoretical arguments by 
    large-scale numerical simulations of charge transport in the 
    extended Fermi-Hubbard chain, where the nearest-neighbor interactions are 
    treated as a perturbation to the integrable reference system.
  \end{abstract}
  
  \maketitle
  
  
  \section{Introduction}
  
  Questions of equilibration and thermalization in isolated 
  quantum systems have experienced a renaissance in recent years 
  \cite{Dalessio2016,Borgonovi2016,Gogolin2016}. However, notwithstanding the 
  significant progress that has been made \cite{Paeckel2019}, describing the 
  precise dynamics of a given quantum many-body system still remains a very 
  challenging task. ``Universal'' principles, which provide a faithful 
  understanding of a wide class of models in various nonequilibrium 
  situations, are therefore highly desirable \cite{Dziarmaga2010,
    Polkovnikov2011,Erne2018,Pruefer2018,Reimann2020,Alhambra2020}. 
  A particularly successful strategy in this context has been the usage of 
  random-matrix ensembles which mimic certain aspects of the full many-body 
  problem \cite{Brody1981}. Prominent examples include the eigenstate 
  thermalization hypothesis \cite{Deutsch1991,Srednicki1994,Rigol2008}, which 
  asserts that the matrix structure of observables becomes essentially random in 
  the eigenbasis of chaotic Hamiltonians \cite{Khaymovich2019,Richter2020c,
    Schonle2020,Brenes2021}, as well as random-circuit models 
  \cite{VonKeyserlingk2018,Nahum2018,Khemani2018}, which have led to new insights 
  into the emergence of hydrodynamics and information scrambling in isolated 
  quantum systems.
  
  A particularly intriguing and omnipresent question in physics is  how the 
  dynamics of a given quantum system is affected by the presence of a 
  perturbation \cite{Deutsch1991,Knipschild2018,Nation2019,
    Dabelow2020,Dabelow2021,Richter2020a}, i.e., scenarios where the Hamiltonian 
  ${\cal H}$ of the full system can be written as
  \begin{equation}\label{Eq::Decomposition}
    {\cal H} = {\cal H}_0 + \varepsilon {\cal V}\ , 
  \end{equation}
  with ${\cal H}_0$ being an unperturbed reference system and $\varepsilon$
  denoting the strength of the perturbation ${\cal V}$. This includes, e.g., the 
  phenomenon of prethermalization \cite{Berges2004,Moeckel2008,Bertini2015,
    Mori2018,Reimann2019a,Mallayya2019}, where ${\cal V}$ weakly breaks a 
  conservation law of the (usually integrable) ${\cal H}_0$, and also the 
  analysis of imperfect echo protocols \cite{Schmitt2018,Dabelow2020a}, where the 
  respective Hamiltonians governing the forward and backward time evolutions are 
  different. In an even broader context, the impact of perturbations also plays 
  an important role for simulations on today's noisy intermediate-scale quantum 
  devices \cite{Preskill2018}, where ${\cal V}$ can be interpreted as the 
  inevitable imperfections of elementary gates which alter the desired circuit 
  \cite{Aruteetal2019}.
  
  Given a quantum system with many degrees of freedom, the impact of a
  perturbation can clearly be manifold. It is therefore quite remarkable that a 
  series of recent works predict that, in the overwhelming majority of cases, the 
  reference dynamics is just damped by a simple function 
  \cite{Dabelow2020,Dabelow2021}, e.g., exponentially as expected from Fermi's 
  golden rule \cite{Genway2013,Richter2020a,Werner2020}. 
  In essence, these works rely on random-matrix theory as ${\cal V}$ is modeled 
  by (an ensemble of) random matrices with respect to the eigenstates of 
  ${\cal H}_0$ \cite{Dabelow2020}, as well as on the concept of typicality 
  \cite{Popescu2006,Goldstein2006,Bartsch2008,Reimann2018}, 
  as a given perturbation is shown to behave very similar to the ensemble 
  average. However, while these predictions were found to compare favorably to a 
  variety of experimental and numerical examples \cite{Dabelow2020}, it yet 
  remains unclear how frequent and under which conditions counterexamples to the 
  typical behavior occur.
  
  In this work, we discuss exactly this question from the perspective of 
  projection-operator techniques, which are well established in the realm of open 
  quantum systems \cite{Breuer2007}. In this way, we provide a fresh insight and 
  show that, within these techniques and under mild assumptions, the ``standard'' 
  case of exponential damping  emerges for the density matrix in the interaction 
  picture but not {\it necessarily} in the Schr\"odinger picture. We particularly 
  suggest that a nontrivial damping in the Schr\"odinger picture can emerge if 
  the dynamics in the unperturbed system possesses rich features. 
  This suggestion has consequences for the time dependence of correlation 
  functions. It is substantiated by large-scale numerical simulations of charge 
  transport in the extended Fermi-Hubbard chain, where the nearest-neighbor 
  interactions are treated as a perturbation to the integrable reference system.
  
  This paper is structured as follows. In  Sec.~\ref{sec:setup} we first establish 
  the setup by introducing an exemplary model 
  and observable, and then turn to a description of our projection-operator 
  approach and its implications on the relaxation dynamics in 
  perturbed many-body quantum systems. We show illustrating numerical 
  results in Sec.~\ref{sec:numerics} and conclude in Sec.~\ref{sec:conclusion}.
  
  
  \section{Setup and projection-operator approach} \label{sec:setup}
  \subsection{Model and observable}\label{sec:setup:model}
  
  Even though our analytical  
  reasoning can be applied to arbitrary operators and Hamiltonians, we here 
  consider for concreteness the dynamics of the particle current in the extended  
  Fermi-Hubbard chain, which constitutes a physically relevant many-body quantum 
  problem (see Refs.~\cite{Zotos2004a,Bertini2020,Bulchandani2021} and references therein). 
  The Hamiltonian of this model reads ${\cal H}= \sum_{r=1}^L
    h_r$ and is a sum over $L$ local terms
  \begin{eqnarray}\label{Eq::Hamiltonian}
    h_r = &-&t_\text{h} \sum_{\sigma = \uparrow, \downarrow} (c_{r,
      \sigma}^\dagger c_{r+1, \sigma}^{\phantom{\dagger}} + \text{H.c.}) \nonumber 
    + U (n_{r, \uparrow} -\tfrac{1}{2})(n_{r, \downarrow} -\tfrac{1}{2}) \\ &+& 
    U' \sum_{\sigma, \sigma'} (n_{r, \sigma} -\tfrac{1}{2})(n_{r+1, \sigma'}
    -\tfrac{1}{2})\, ,
  \end{eqnarray}
  where we assume periodic boundary conditions, i.e., we have $L+1 \equiv 1$.  
  $c_{r, \sigma}^\dagger$ ($c_{r, \sigma}^{\phantom{\dagger}}$) creates (annihilates) 
  a fermion with spin $\sigma$ at lattice site $r$ and $n_{r,\sigma} = c_{r,
        \sigma}^\dagger c_{r, \sigma}^{\phantom{\dagger}}$ is the occupation-number 
  operator. $t_\text{h}$ is the hopping matrix element and $U, U' > 0$ denote the 
  strengths of the repulsive on-site and nearest-neighbor interactions, 
  respectively. While the model is noninteracting for $U, U' = 0$, it in fact 
  remains integrable in terms of the Bethe ansatz also for finite on-site 
  interactions $U > 0$ \cite{Essler2005}. In contrast, this integrability is 
  broken for any $U' > 0$. Note that ${\cal H}$ preserves the number of each 
  fermion species. 
  
  As an observable, we consider the particle current. It can be derived from a  
  continuity equation and takes on the well-known form (see Refs.~\cite{Zotos2004a,Bertini2020,Bulchandani2021} and references therein)
  ${\cal J} = \sum_{r=1}^L j_r$,
  \begin{equation}
    j_r = -t_\text{h} \sum_{\sigma = \uparrow, \downarrow} [(i \, c_{r,
          \sigma}^\dagger c_{r+1, \sigma}^{\phantom{\dagger}} + \text{H.c.})] \, .
  \end{equation}
  While the particle current does not depend on $U$ and $U'$, its dynamics does. 
  Only in the case $U = U' = 0$, we have $[ {\cal J}, {\cal H}] = 0$. Generally,
  $\text{tr}[ {\cal J} ] = 0$ and $\text{tr}[ {\cal J}^2 ] = D L t_\text{h}^2/4$, 
  where $D = 4^L$ is the dimension of the Hilbert space. In
  this paper, we will be particularly concerned with the dynamics 
  of current-current correlation functions. However, as already stated above, 
  all that follows now carries over to other choices of 
  observable and Hamiltonian.

  \subsection{Projection-operator approach}\label{sec:setup:po}
  
  To apply projection-operator techniques, we first 
  decompose the full system $\cal H$ according to Eq.~\eqref{Eq::Decomposition}
  into an unperturbed system ${\cal H}_0$ and a perturbation $\varepsilon \, \cal
    V$. For instance, for the Fermi-Hubbard chain \eqref{Eq::Hamiltonian}, we will 
  later consider two different reference systems ${\cal H}_0$, i.e., the 
  noninteracting ${\cal H}_0 = {\cal H}(U = U' = 0)$ and the interacting 
  integrable ${\cal H}_0 = {\cal H}(U \neq 0, U' = 0)$.
  
  After this decomposition, we then define a projection superoperator ${\cal
        P}$, which projects a density matrix $\rho(t)$ at time $t$ onto a set of 
  relevant degrees of freedom. This set should at least include the identity and 
  the observable of interest,
  \begin{equation}
    {\cal P} \, \rho(t) = \frac{1}{D} + \frac{C(t)}{\langle {\cal
        J}^2 \rangle} \, {\cal J} \, , \quad C(t) = \langle {\cal J} \rho(t)
    \rangle \, , \label{projection}
  \end{equation}
  where $\langle \bullet \rangle = \text{tr}[\bullet]/D$. 
  Due to $\langle {\cal J} \rangle = 0$, ${\cal P}^2 = {\cal P}$. 
  In this work, we are interested in the time-dependent part
  $C(t)$ of the projected density matrix. It is important to 
  note that, using the projection in Eq.~(\ref{projection}), $C(t)$ is not 
  identical in the Schr\"odinger and interaction picture. Even though $C(t)$ is 
  a coefficient and not an expectation value, it will turn out below that $C(t)$
  can be expressed in terms of certain types of correlation 
  functions. Importantly, throughout this paper, the notions of 
  Schr\"odinger or interaction picture should be understood with respect to 
  the dynamics 
  of the density matrix. This wording should not be confused 
  with the fact that expectation values of observables 
  are the same in both pictures.
  
  While taking into account more degrees of freedom is possible, this will not be 
  necessary for our purposes. In particular, for initial conditions $\rho(0)$ in 
  the span of $1$ and $\cal J$, we further have ${\cal P} \rho(0) = \rho(0)$. 
  From now on, we will focus on such kind of initial conditions, which also 
  appear in the context of linear response theory \cite{LRT}
  \nocite{Kubo1991,Brenig1989}.
  
  After having defined the projection superoperator (and the reference system), 
  the so-called time-convolutionless (TCL) projection-operator technique 
  routinely leads to a time-local differential equation for the evolution of 
  ${\cal P} \rho_\text{I}(t)$ in the interaction picture \cite{Breuer2007,
    chaturvedi1979},
  \begin{equation}\label{Eq::TCL}
    \frac {\partial}{\partial t} {\cal P} \rho_\text{I}(t) = {\cal G}(t) \, {\cal
        P} \rho_\text{I}(t) + {\cal I}(t) \, (1 - {\cal P}) \, \rho(0) \, ,
  \end{equation}
  where $\rho_\text{I}(t) = e^{i {\cal H}_0 t} e^{-i {\cal H} t}
    \rho(0) e^{i {\cal H} t} e^{-i {\cal H}_0 t}$. The term ${\cal I}(t)$, i.e., 
  the inhomogeneity on the right hand side (r.h.s.) of Eq.\ \eqref{Eq::TCL}, 
  can be neglected, due to $(1 - {\cal P}) \, \rho(0) = 0$. 
  The generator ${\cal G}(t)$ is given as a systematic 
  series expansion in powers of $\varepsilon$. In many cases, just like in our 
  case, odd orders vanish. Hence, the lowest order is the second order and reads
  \begin{equation}
    {\cal G}_2(t) = \varepsilon^2 \! \int_0^t \! \text{d}t' \, {\cal P} {\cal L}(t) 
    {\cal
        L}(t') {\cal P} \, ,
  \end{equation}
  where the Liouvillian is given by ${\cal L}(t) \, \bullet = -i [{\cal
        V}_\text{I}(t), \bullet]$ with
  ${\cal V}_\text{I}(t) = e^{i {\cal H}_0 t} \, {\cal V}(0) \, e^{-i
    {\cal H}_0 t}$.
  
  So far, we have invoked no significant assumption. The central assumption in 
  the following will be a truncation to lowest order. The quality of such a 
  truncation naturally depends on the perturbation strength $\varepsilon$, but 
  also on the structure of the perturbation $\cal V$ and the degrees of 
  freedom in the projection superoperator $\cal P$. Note,
  however, that a truncation to lowest order does not necessarily imply that we have 
  to restrict ourselves to weak perturbations $\varepsilon\rightarrow0$
  and long times $t\rightarrow\infty$. In particular, a lowest-order 
  truncation turns out to be reasonable in many situations \cite{Bartsch2008}. 
  While the quality can naturally be 
  further improved by taking into account higher-order corrections 
  \cite{Steinigeweg2011b}, conditions for neglecting 
  higher orders at rather large $\varepsilon$ can be found in
  Ref.~\cite{Steinigeweg2011b}.
  
  Now, we use a truncation to lowest order, as well as the simple
  mathematical facts that,
  \begin{equation}
    C(t) \propto \langle {\cal J}(t) {\cal J}\rangle\ ,\quad
    C_\text{I}(t)\propto \langle {\cal
      J}(t) {\cal
      J}_\text{I}(t) \rangle\ , 
  \end{equation}
  which relate the time-dependent part of the density
  matrix (in the Schr\"odinger or interaction picture) to a certain type of  
  correlation function. (A derivation of this relation can be
  found in Appendix\ \ref{app:proof_correlationfunc}.) In
  particular, within the TCL formalism, a rate 
  equation can be obtained for $C_\text{I}(t)$ in the interaction picture,
  \begin{equation}
    \frac {\partial}{\partial t} \langle {\cal J}(t) {\cal J}_\text{I}(t)
    \rangle = -\varepsilon^2 \, \gamma_2(t) \, \langle {\cal J}(t) {\cal
      J}_\text{I}(t)
    \rangle \, , \label{rate_equation}
  \end{equation}
  where ${\cal J}(t) = e^{i {\cal H} t} \, {\cal J}(0) \, e^{-i
    {\cal H} t}$ and the time-dependent damping $\gamma_2(t)$ results from  a 
  time-integral over a kernel $k_2(t,t') = k_2(\tau = t - t')$,
  \begin{equation}
    \gamma_2(t) = \int_0^t \!\! \text{d}\tau \, k_2(\tau) \, , \, 
    k_2(\tau) = 
    \frac{\langle i [{\cal J}, {\cal V}_\text{I}(\tau)]i[{\cal J}, {\cal
        V}_\text{I}] \rangle}{\langle \mathcal{J}^2 \rangle} \, . \label{kernel_rate}
  \end{equation}
  Apparently, if $k_2(\tau) \to 0$ for sufficiently long 
  times, then we have $\gamma_2(t) \to \text{const.}$ at such time scales. We 
  note that the kernel $k_2(\tau)$ can in principle be 
  calculated analytically in the thermodynamic limit, if the reference system 
  ${\cal H}_0$ is integrable. But often a numerical calculation of 
  $k_2(\tau)$ in systems of finite size is sufficient 
  \cite{Steinigeweg2011b}.
  
  The solution of rate equation (\ref{rate_equation}) obviously is an
  exponential decay of the form
  \begin{equation}
    \frac{\langle {\cal J}(t) {\cal J}_\text{I}(t) \rangle}{\langle {\cal J}^2
      \rangle} = \exp \Big [ - \varepsilon^2 \! \int_0^t \! \text{d}t' \,
      \gamma_2(t') \Big ] \, . \label{exponential_prediction}
  \end{equation}
  This solution reflects our central result: Within our TCL approach, 
  the time-dependent part of the density matrix is damped 
  exponentially in the interaction picture and not {\it necessarily} in the 
  Schr\"odinger picture. Clearly, both pictures must agree, if the observable 
  $\cal J$ is preserved in the reference system ${\cal H}_0$, $[{\cal J}, {\cal
            H}_0] = 0$. For instance, for the particle current in the Fermi-Hubbard chain, 
  this preservation is given in the noninteracting ${\cal H}_0 = {\cal H}(U = U'
    = 0)$. Therefore, both pictures can also be expected to be rather similar, 
  whenever the dynamics of ${\cal J}_\text{I}(t)$ is sufficiently slow compared 
  to the dynamics of ${\cal J}(t)$. In the general situation, however, the two 
  pictures are just not the same:
  \begin{equation}
    \frac{\langle {\cal J}(t) {\cal J} \rangle_{\varepsilon > 0}}{\langle {\cal
        J}(t) {\cal J} \rangle_{\varepsilon = 0}} \neq \frac{\langle {\cal J}(t) {\cal
        J}_\text{I}(t) \rangle}{\langle {\cal J}^2
      \rangle} \, .
  \end{equation}
  Hence, {\it a priori}, one cannot expect that the lowest-order prediction of an 
  exponential decay in Eq.~(\ref{exponential_prediction}) simply carries over to 
  the Schr\"odinger picture, and the relaxation dynamics
  of $\langle {\cal J}(t) {\cal J}\rangle_{\varepsilon>0}$ may exhibit nontrivial 
  behavior that is distinct from typicality predictions in 
  Refs.~\cite{Dabelow2020,Dabelow2021}. For 
  instance, as we demonstrate later, this difference is eye striking for the 
  strongly interacting reference system ${\cal H}_0 = {\cal H}(U \gg t_\text{h},
    U' = 0)$. Note that Eq.~(\ref{exponential_prediction}) has also consequences 
  for transport quantities.
  
  
  \section{Numerical illustration}\label{sec:numerics}
  
  Next, we illustrate our central result in 
  numerical simulations after a description of the employed method.
  
  \subsection{Method}
  
  To study system sizes larger than what 
  is possible with full exact diagonalization (ED), we rely on the concept of 
  dynamical quantum typicality (DQT) \cite{Popescu2006,Goldstein2006,
    Bartsch2009, Reimann2018} and obtain time-dependent autocorrelation 
  functions from a single pure state $| \psi \rangle$, which is drawn at random 
  from a high-dimensional Hilbert space. While this approach is by now well 
  established for ``standard'' correlation functions such as
  $\langle {\cal J}(t){\cal J}\rangle$
  (see Refs.~\cite{Jin2021,Heitmann2020} and references therein), 
  the dynamics of correlation functions with a more complicated
  time dependence such as $\langle {\cal J}(t) {\cal J}_\text{I}(t)\rangle$ can 
  be obtained in a rather similar 
  fashion. Specifically, we first introduce the two auxiliary pure states
  \begin{eqnarray}
    | \phi (t) \rangle &=&  e^{-i {\cal H} t} \, e^{i {\cal H}_0 t} \, | \psi 
    \rangle \, , \label{state1} \\
    | \varphi (t) \rangle &=&  e^{-i {\cal H} t} \, e^{i {\cal H}_0 t} \, {\cal J}
    \, | \psi \rangle 
    \, ,
  \end{eqnarray}
  and then approximate the autocorrelation function and its time dependence as
  \begin{equation}
    \langle {\cal J}(t) {\cal J}_\text{I}(t) \rangle = \frac{\langle \phi(t) | 
    {\cal J}| \varphi(t) \rangle}{\langle \phi | \phi \rangle} + {\cal 
    O} \Big (\frac{1}{\sqrt{D}} \Big ) \, ,
  \end{equation}
  where the statistical error becomes negligibly small for 
  system sizes studied here. Compared to the usual approximation 
  of $\langle {\cal J}(t){\cal J}\rangle$
  \cite{Iitaka2003,Elsayed2013}, the 
  approximation of $\langle {\cal J}(t) {\cal J}_\text{I}(t)\rangle$ is more 
  costly from a numerical point 
  of view, since at each point in time an additional backward propagation
  with respect to the reference system ${\cal H}_0$ is required. However, 
  this extra operation can still be carried out in large Hilbert spaces beyond the 
  range of full ED, thereby reducing the impact of finite-size effects,
  see Appendix\ \ref{app:finitesizescaling} for a detailed analysis of finite-size effects.
  In this paper, we treat systems with up to $L = 16$, where $D \approx
    4.3 \times 10^{9}$, and the largest symmetry subspace has a dimension 
  $\approx 10^7$.
  Note that a time evolution of the form (\ref{state1}) is also 
  relevant for the stability of quantum motion with respect to a static 
  perturbation \cite{Prosen2002}.
  
  \subsection{Results}\label{sec:results}
  
  Let us now turn to our actual numerical results for the two scenarios of a noninteracting and an interacting reference system.
  \subsubsection{Noninteracting reference system}\label{sec:results:nonint}
  \begin{figure}[t]
    \centering
    \includegraphics[width=.45\textwidth]{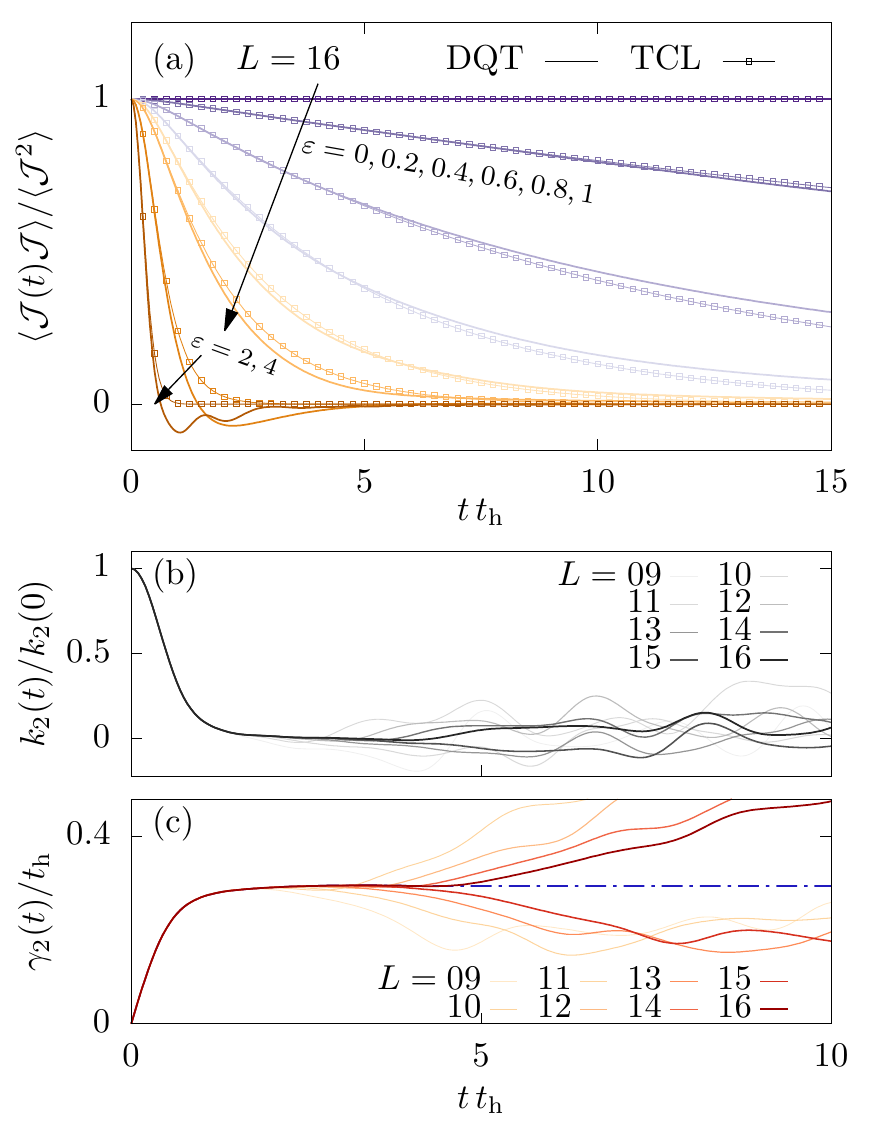}
    \caption{(a) Decay of the current autocorrelation function 
    $\langle \mathcal{J}(t) \mathcal{J} \rangle$  for the noninteracting ${\cal
          H}_0 = {\cal H}(U = U' = 0)$, which is perturbed by particle-particle 
    interactions of various strengths $U/t_\text{h} = U'/t_\text{h} = \varepsilon
      \leq 4 $. Numerical results from DQT in a finite system of size $L = 16$
    are compared to the lowest-order prediction of the TCL projection-operator 
    technique in Eq.~(\ref{exponential_prediction}). (b) Kernel $k_2(t)$ and (c) 
    rate $\gamma_2(t)$, as both given in Eq.~(\ref{kernel_rate}), for $L
      \leq 16$. Finite-size effects are mild and systematic. The expected plateau 
    value of $\gamma_2(t)$ for $L \to \infty$ is indicated (dash-dotted line).}
    \label{fig:1}
  \end{figure}
  We start with a decomposition where 
  the reference system is chosen to be noninteracting, ${\cal H}_0 = {\cal H}(U =
    U' = 0)$, such that Schr\"odinger and interaction picture are identical, 
  $C(t) = C_\text{I}(t)$, due to 
  $[{\cal J}, {\cal H}_0] = 0$. The role of the perturbation is then played by 
  the particle-particle interaction terms. In Fig.~\ref{fig:1}, we summarize the 
  decay of the current autocorrelation function $\langle {\cal J}(t) {\cal J}
    \rangle = \langle {\cal J}(t) {\cal J}_\text{I}(t) \rangle$ for a finite system 
  of size $L = 16$ and interaction strengths  $U/t_\text{h} = U'/t_\text{h} =
    \varepsilon \leq 4$. The decay is faster the larger $\varepsilon$, and an 
  exponential relaxation for weak $\varepsilon$ changes into a Gaussian type of 
  relaxation for stronger $\varepsilon$. This 
  overall behavior is in qualitative agreement with the lowest-order prediction 
  of the TCL projection-operator technique in Eq.~(\ref{exponential_prediction}). 
  Note that the Gaussian behavior is expected due 
  to $\gamma_2(t) \propto t$ at small $t$, which become relevant for large 
  $\varepsilon$ \cite{Steinigeweg2011b}.
  
  To exclude that this agreement is accidental, we depict in Figs.~\ref{fig:1}(b) 
  and \ref{fig:1}(c) numerical results for the kernel $k_2(t)$ and 
  rate $\gamma_2(t)$, as both given in Eq.~(\ref{kernel_rate}), for different 
  chain lengths $L \leq 16$. Apparently, $k_2(t)$ decays fast to zero, and the 
  visible finite-size effects set in at time scales after this initial decay. As 
  a consequence, the damping $\gamma_2(t)$ shows a mild dependence on system size 
  and, in particular, a conclusion on the plateau value of $\gamma_2(t)$ for $L
    \to \infty$ is possible. Therefore, we can quantitatively evaluate the 
  lowest-order prediction in Eq.~(\ref{exponential_prediction}) and compare to 
  the direct numerics in Fig.~\ref{fig:1} discussed before. We find that the 
  agreement is remarkably good over a wide range of perturbation strengths 
  $\varepsilon$, and small differences might be either related to residual 
  finite-size effects or missing higher-order corrections.  
  A detailed analysis of finite-size effects can be found in
  Appendix\ \ref{app:finitesizescaling}.
  Thus, our TCL approach correctly captures the 
  ``standard'' case of exponential damping of $\langle {\cal
      J}(t){\cal J}\rangle$, in agreement with typicality and random-matrix 
  considerations \cite{Dabelow2020, Dabelow2021,
    Richter2020a}.
  
  \subsubsection{Interacting reference system}\label{sec:results:int}
  
  \begin{figure}[t]
    \centering
    \includegraphics[width=.45\textwidth]{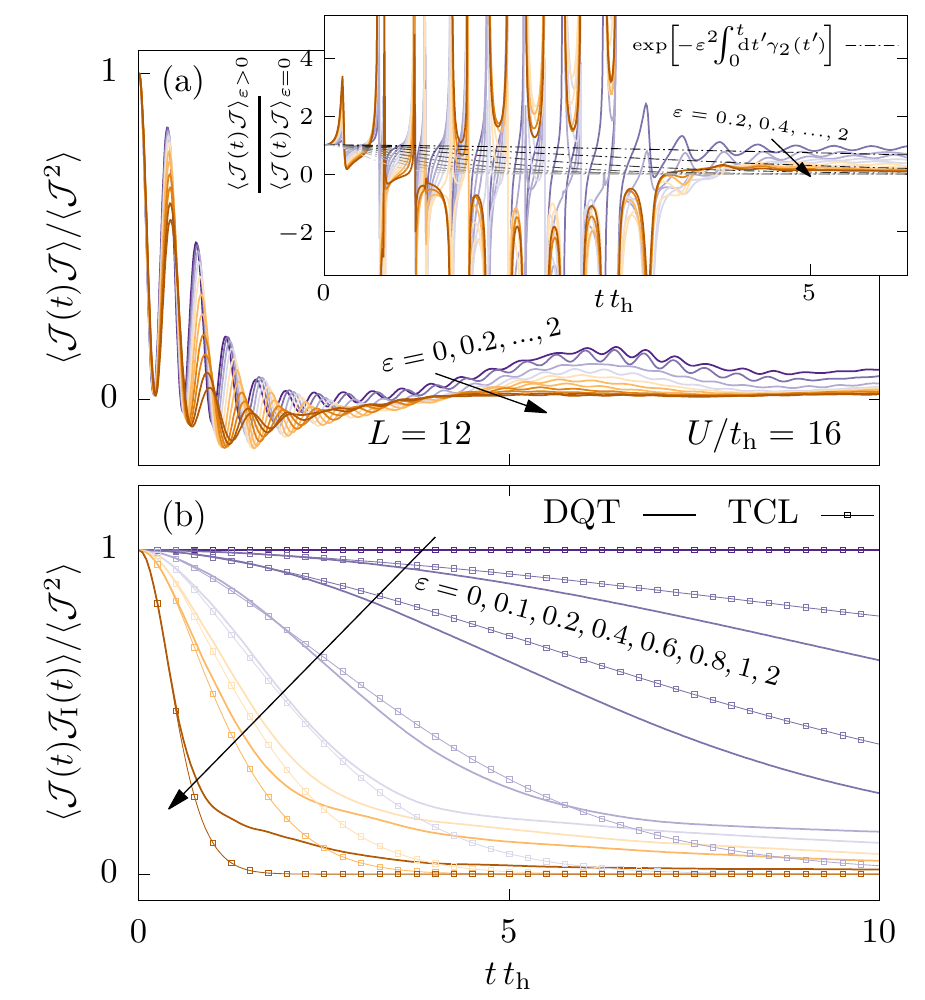}
    \caption{Decay of (a) 
    $C(t) = \langle \mathcal{J}(t) \mathcal{J} \rangle$ and 
    (b) $C_\text{I}(t) = \langle \mathcal{J}(t) \mathcal{J}_{\mathrm{I}}(t)
      \rangle$ for the interacting $\mathcal{H}_0 =
      \mathcal{H}(U/t_\text{h} = 16, U' = 0)$, which is perturbed in this case by 
    nearest-neighbor particle-particle interactions of strength $U'/t_\text{h} =
      \varepsilon \leq 2$. Numerical results from DQT are shown for a finite 
    system of size $L = 12$ and in (b) additionally compared to the lowest-order 
    prediction of the TCL projection-operator technique in Eq.~(\ref{exponential_prediction}). 
    For the kernel $k_2(t)$ and rate $\gamma_2(t)$, 
    see \figref{app-fig:2} in Appendix\ \ref{app:finitesizescaling}. 
    The inset in (a) shows the ratio between $\langle {\cal
        J}(t) {\cal J} \rangle_{\varepsilon > 0}$ (perturbed dynamics) and $\langle  
      {\cal J}(t) {\cal J} \rangle_{\varepsilon = 0}$ (unperturbed dynamics) in the 
    Schr\"odinger picture. This ratio is a nontrivial function and does not 
    coincide with the damping in the interaction picture.}
    \label{fig:3}
  \end{figure}
  Finally, and most importantly in the context of this paper, we turn to the 
  decomposition ${\cal H} = {\cal H}_0 + \varepsilon {\cal V}$  with 
  an interacting reference system, $\mathcal{H}_0 = \mathcal{H}(U \neq 0, U' =
    0)$, where the Schr\"odinger and interaction picture are no longer the same, 
  $C(t)\neq C_I(t)$, and 
  the perturbation ${\cal V}$ is given by the nearest-neighbor interaction 
  $U'>0$. We decide to choose a large on-site interaction $U/t_\text{h} = 16 \gg
    1$, since the dynamics for such $U$ is known to have rich features 
  \cite{Karrasch2014a, Jin2015}. However, 
  the overall phenomenology emerges for smaller values of $U$ as well,
  as can be seen in the additional data presented in Appendix\ \ref{app:smallU}.
  As shown in Fig.~\ref{fig:3}(a) for a finite system size $L = 12$, 
  $C(t) = \langle {\cal J}(t) {\cal J} \rangle$ in the 
  Schr\"odinger picture exhibits oscillatory behavior, where the frequencies and 
  zero crossings also vary with the strength $U'/t_\text{h} = \varepsilon$ of the 
  nearest-neighbor interaction. Hence, from visual inspection, it is clear  
  that unperturbed and perturbed dynamics cannot be related by a simple damping 
  function. Their nontrivial relation becomes even more 
  obvious by plotting their ratio (see the inset of Fig.~\ref{fig:3}).
  
  In the interaction picture, however, the situation turns out to be different. 
  As shown in Fig.~\ref{fig:3}(b), the behavior of 
  $C_\text{I}(t) = \langle {\cal J}(t) {\cal J}_\text{I}(t)\rangle$ is 
  like the one in Fig.~\ref{fig:1}. It decays monotonously and changes from 
  exponential to Gaussian type of relaxation as $\varepsilon$ is increased, in 
  qualitative agreement with the lowest-order prediction of the TCL 
  projection-operator technique in Eq.~(\ref{exponential_prediction}). 
  Furthermore, a quantitative comparison is also feasible, since the 
  corresponding kernel $k_2(t)$ and rate $\gamma_2(t)$ are converged with respect to 
  system size (see \figref{app-fig:4} in Appendix\
  \ref{app:finitesizescaling}), 
  at least for the time scales depicted in Fig.~\ref{fig:3}(b). 
  Apparently, the agreement is not as convincing as before and 
  deviations set in for times $t \, t_\text{h} \sim 4$. However, for such times, 
  the direct numerics is known to still depend on system size 
  (see, e.g., Refs.~\cite{Karrasch2014a,Jin2015} and 
  Appendix\ \ref{app:finitesizescaling}), 
  and deviations might eventually disappear in the 
  thermodynamic limit $L \to \infty$. We should also stress 
  that the restriction by the finite-size time $t \, t_\text{h} \sim 4$ does not 
  allow us to study very weak perturbations $\varepsilon \ll 0.1$ in our 
  numerical simulation, where the relaxation takes place on a much longer time 
  scale. How to numerically study the limit of very weak $\varepsilon$
  therefore remains an open problem.
  
  In the specific context of currents, this result also has direct consequences 
  for the transport behavior \cite{Zotos2004, Jung2006, Steinigeweg2016}; i.e., 
  only in the interaction picture is the dynamics of the density
  matrix exponentially damped due to perturbations such that (i) the 
  frequency dependence of the 
  conductivity has a simple Lorentzian form and (ii) the dc conductivity 
  $\sigma_\text{dc}$ scales as $\sigma_\text{dc} \propto 1/\varepsilon^2$. But 
  for the dynamics of the density matrix in
  the Schr\"odinger picture, which is of actual interest, both (i) and (ii) 
  cannot be 
  expected (see also the corresponding data shown in
  Appendix\ \ref{app:conductivity}).
  
  \section{Conclusion} \label{sec:conclusion}
  
  We have addressed the question of how the dynamics of a given 
  quantum system is altered when a perturbation is switched on. 
  We have shown that, within our analytical approach based 
  on projection-operator techniques, the ``standard'' case of exponential damping 
  occurs for the density matrix in the interaction picture but not {\it necessarily} in the 
  Schr\"odinger picture. This key point we have illustrated explicitly in 
  numerical simulations for charge transport in the strongly interacting extended 
  Fermi-Hubbard chain, as a physically relevant many-body problem. Using this 
  example, we have unveiled the emergence of nontrivial damping 
  of current-current correlation functions, which is on the one 
  hand not expected from typicality and random-matrix 
  considerations and on the other hand demonstrates the complexity of quantum 
  many-body systems out of equilibrium. While our numerics has focused on one 
  specific example, our analytical reasoning suggests a similar behavior for other 
  quantum systems. We expect that a nontrivial damping of 
  relaxation dynamics in perturbed many-body quantum systems
  occurs most likely for cases where already the unperturbed dynamics 
  possesses rich features. Thus, strongly interacting spin-$1/2$ XXZ chains or 
  ladders \cite{Karrasch2014, Karrasch2015, Steinigeweg2014b} are natural 
  candidates and promising future directions of research. 
  However, we do not expect that integrability is a necessary 
  condition. It would also be desirable to tackle the open 
  problem of how to numerically study the limit of very weak perturbations.
  
  
  \section*{Acknowledgments}
  
  We sincerely thank L. Dabelow and P. Reimann for 
  fruitful discussions. This work has been funded by the Deutsche 
  Forschungsgemeinschaft (DFG), Grants No.~397107022 (GE 1657/3-2), No.~397067869 (STE 2243/3-2), within the DFG Research Unit FOR 2692, Grant No.~355031190. 
  J.~R. has been funded by the European Research 
  Council (ERC) under the European Union's Horizon 2020 research and innovation 
  programme (Grant agreement No.~853368).
  
  \begin{appendix}

    \section{Other values for the on-site interaction}\label{app:smallU}
    
    \begin{figure}[t]
      \centering
      \includegraphics[width=.49\textwidth]{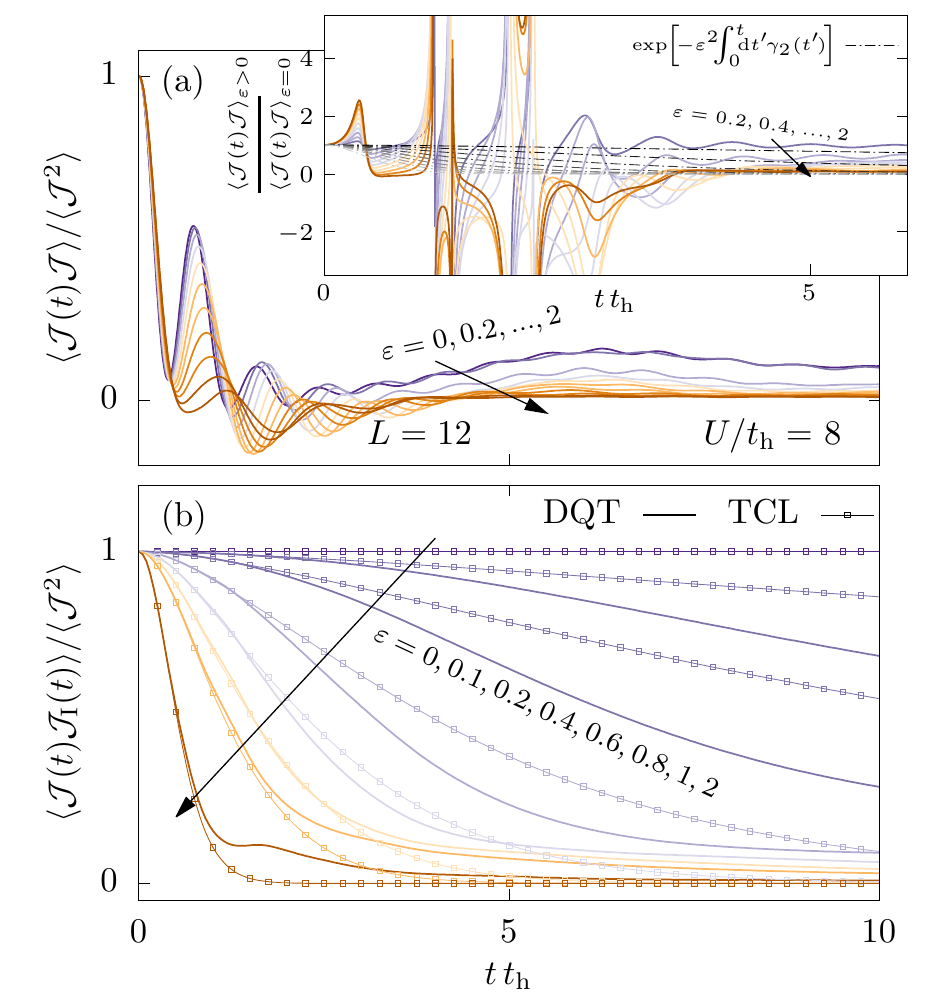}
      \caption{Similar data as depicted in Fig.~\ref{fig:3}, but now shown for the on-site interaction strength $U/t_\mathrm{h} = 8$.}
      \label{app-fig:1}
    \end{figure}
    \begin{figure*}[t]
      \centering
      \includegraphics[width=.92\textwidth]{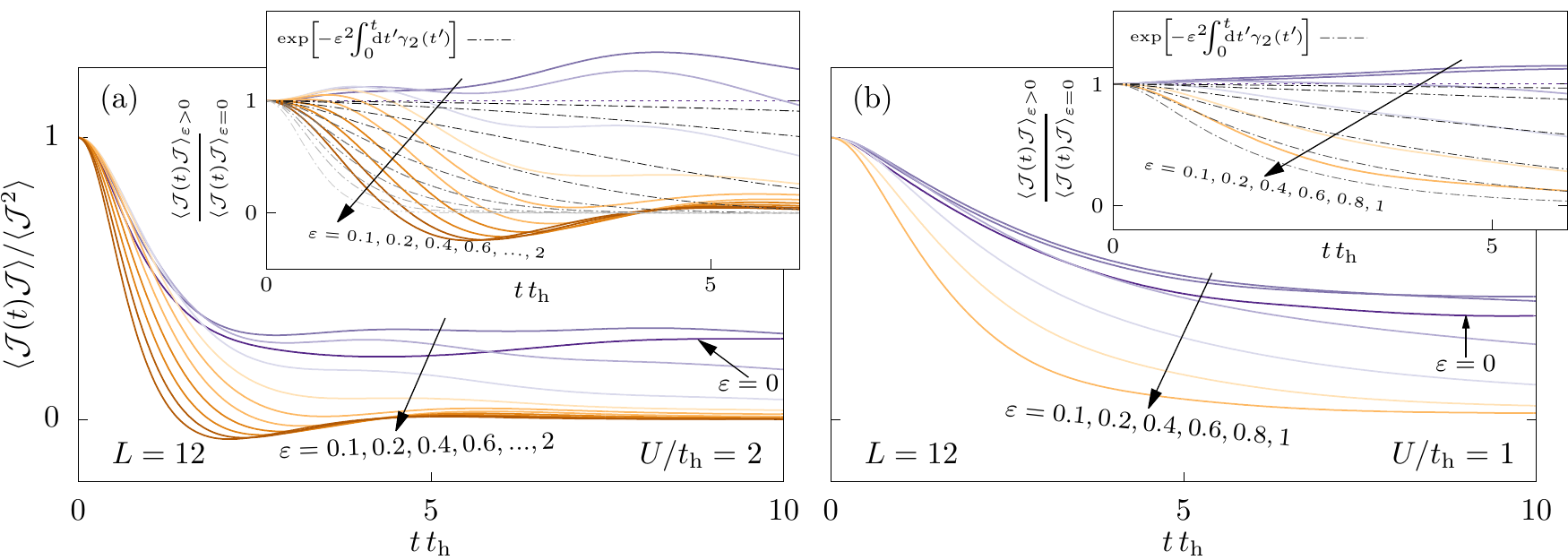}
      \caption{Similar data as depicted in \figref{app-fig:1}(a), but now shown for smaller on-site interaction strengths (a) $U/t_\mathrm{h} = 2$ and (b) $U/t_\mathrm{h} = 1$.}
      \label{app-fig:1b}
    \end{figure*}
    
    Since we have focused in the main text on a strongly interacting reference 
    system ${\cal H}_0 = {\cal H}(U/t_\mathrm{h} \gg 1, U' = 0)$ with a single 
    on-site interaction strength $U/t_\mathrm{h} = 16$, we redo the calculation in 
    Fig.~\ref{fig:3} for another value of $U$. As shown in Fig.~\ref{app-fig:1}, the 
    overall picture remains the same for $U/t_\mathrm{h} = 8$. Therefore, our 
    numerical illustration is not fine-tuned with respect to $U$. Additionally, \figref{app-fig:1b} shows similar data as \figref{app-fig:1}(a) but for small interaction strengths $U/t_\mathrm{h} = 2$ and $U/t_\mathrm{h} = 1$. Naturally, with decreasing interaction strength $U$, we start to approach the noninteracting limit where $\cal{J}$ is conserved and the oscillations in the reference dynamics disappear. However, while the ratio $\langle {\cal J}(t) {\cal J} \rangle_{\varepsilon > 0}/\langle {\cal J}(t) {\cal J} \rangle_{\varepsilon = 0}$ assumes a more well-behaved shape, the relation between perturbed and unperturbed dynamics remains hardly reconcilable with exponential damping for $U/t_\mathrm{h} = 2$.
    
    \section{Finite-size scaling}\label{app:finitesizescaling}
    
    \subsection{Interacting reference system}
    
    \begin{figure}[b]
      \centering
      \includegraphics[width=.5\textwidth]{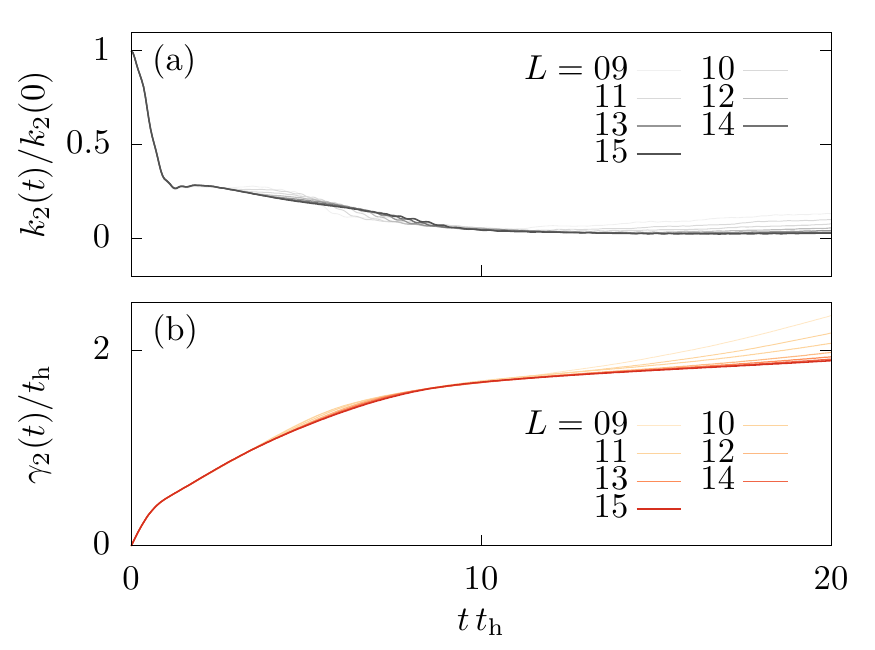}
      \caption{Second-order (a) kernel $k_2(t)$ and (b) damping
        $\gamma_2(t)$, as both given in Eq.\ \eqref{kernel_rate} of the main text, for the interacting 
        $\mathcal{H}_0 = \mathcal{H}(U/t_\text{h} = 16, U' = 0)$. Numerical results from 
        DQT are shown for various chain lengths $L \leq 15$.}
      \label{app-fig:2}
    \end{figure}
    
    \begin{figure}[b]
      \centering
      \includegraphics[width=.5\textwidth]{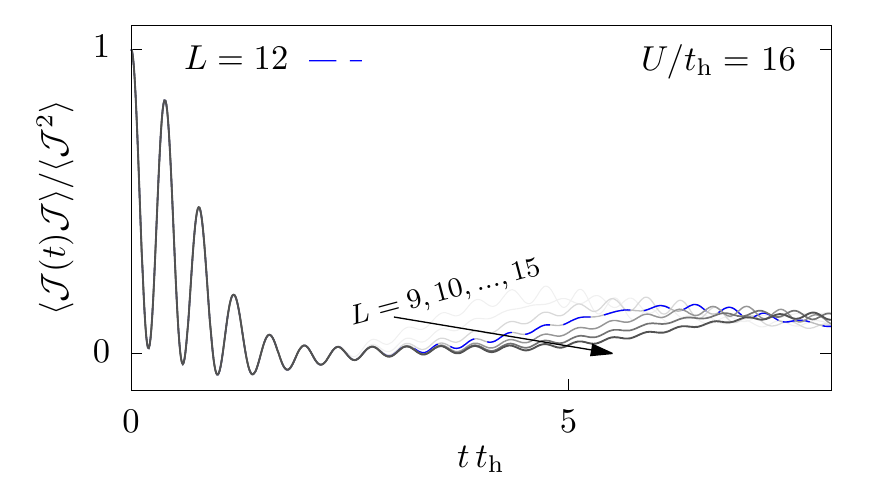}
      \caption{Time dependence of the current autocorrelation function
        $\langle  {\cal J}(t) {\cal J} \rangle$ in the strongly interacting system 
        ${\cal H}(U/t_\mathrm{h} = 16, U' = 0)$, as obtained from DQT for different 
        system sizes $L \leq 15$. For such $L$, data are converged up to times $t \,
          t_\mathrm{h} \sim 4$. Similar data can be found in \cite{Jin2015}.}
      \label{app-fig:3}
    \end{figure}
    
    In the main text, we have mentioned in the context of Fig.~\ref{fig:3} that the corresponding 
    kernel $k_2(t)$ and the rate $\gamma_2(t)$ are converged with respect to system size for the relevant time scales.
    To support this, we depict in Fig.~\ref{app-fig:2} the numerical results for 
    $k_2(t)$ and $\gamma_2(t)$ for different system sizes $L\leq15$.
    We have also mentioned that finite-size effects for the strongly interacting 
    case $U/t_\mathrm{h} = 16$ in Fig.~\ref{fig:3}  occur for times 
    $t \, t_\mathrm{h} \sim 4$. Since we have shown curves for a single system size 
    $L = 12$ there,  we now illustrate in Fig.~\ref{app-fig:3} these finite-size 
    effects explicitly by depicting curves for different system sizes $L \leq 15$. 
    We do so for ${\cal H}(U/t_\mathrm{h} = 16, U' = 0)$, which enters as ${\cal
          H}_0$ the interaction picture for all perturbations $U'/\thubbard = \varepsilon > 0$.
    
    \subsection{Noninteracting reference system}
    
    \begin{figure*}[t]
      \centering
      \includegraphics[width=\textwidth]{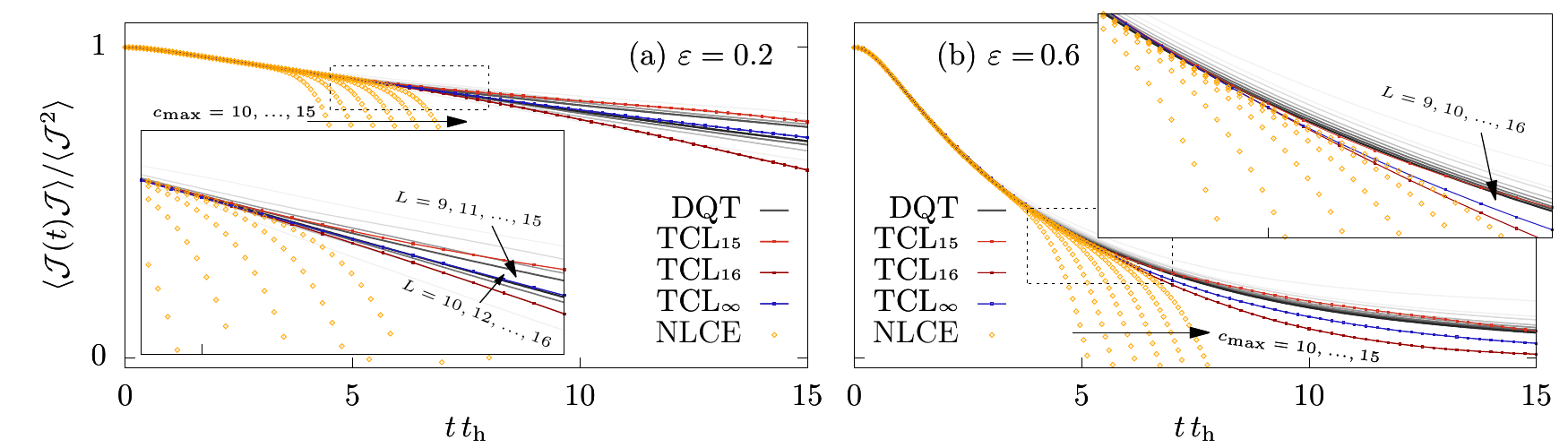}
      \caption{Time dependence of the current autocorrelation function
        $\langle\mathcal{J}(t)\mathcal{J}\rangle$ in the noninteracting system 
        $\mathcal{H}_0=\mathcal{H}(U=U'=0)$, which is perturbed by interactions 
        $U/\thubbard=U'/\thubbard=\varepsilon$ for two exemplary strengths (a) 
        $\varepsilon=0.2$ and (b) $\varepsilon=0.6$. Numerical results from DQT are 
        shown for different system sizes $L=9,...,16$ and compared to the lowest-order 
        prediction of the TCL projection-operator technique in Eq.~\eqref{exponential_prediction}
        of the main 
        text. For the TCL curves, different rates $\gamma_2(t)$ corresponding to finite 
        system sizes $L=15$ (TCL{\tiny15}), $L=16$ (TCL{\tiny16}), as well as its 
        estimate for $L\to\infty$ (TCL{\tiny$\infty$}) are used [cf. Fig.~\ref{fig:1}(c) of
            the main text]. Additionally, NLCE data is shown for expansion orders 
        $c_{\mathrm{max}}\leq15$.}
      \label{app-fig:4}
    \end{figure*}
    
    In the discussion of Fig.~\ref{fig:1} in the main text, we have mentioned that small 
    differences between direct DQT calculations and lowest-order TCL predictions 
    of the current autocorrelation function $\langle\mathcal{J}(t)\mathcal{J}
      \rangle$ might in part be related to residual finite-size effects. To support 
    this, we depict similar data for different system sizes and for two exemplary 
    perturbation strengths $\varepsilon=0.2$ and $\varepsilon=0.6$ in 
    \figref{app-fig:4}. Numerical results from DQT for different system sizes 
    $L\leq16$ are shown and compared to the lowest-order prediction of the TCL 
    projection operator technique in Eq.~\eqref{exponential_prediction} of the main text.
    Here, we show three different curves (labeled as TCL{\tiny 15}, TCL{\tiny 16}, 
    and TCL{\tiny$\infty$}), corresponding to the rate $\gamma_2(t)$ obtained for 
    the two largest numerically accessible chain lengths $L=15,16$ as well as its 
    estimate for the thermodynamic limit $L\to\infty$, featuring a constant plateau 
    for $t\,\thubbard\gtrsim 4$ [see Fig.~\ref{fig:1}(c) of the main text].
    For $\varepsilon=0.2$ [\figref{app-fig:4}(a)], the DQT curves for the 
    largest system sizes are converged up to times $t\,\thubbard \sim 5$ and 
    coincide with all three TCL curves. For later times, both the DQT and the 
    TCL results show mild finite-size effects, whereby the TCL{\tiny$\infty$}
    prediction appears to agree best with the scaling behavior of the DQT 
    data. For $\varepsilon=0.6$ [\figref{app-fig:4}(b)], a very similar 
    behavior is found in the comparison of the DQT and the TCL curves. 
    
    Complementary to the DQT and the TCL data, \figref{app-fig:4} also shows 
    numerical results for $\langle\mathcal{J}(t)\mathcal{J}\rangle$ in the 
    thermodynamic limit $L\to\infty$ as obtained by means of a numerical 
    linked-cluster expansion (NLCE) (see, e.g., Refs.~\cite{Rigol2006,Richter2019b} 
    and below) for different expansion orders $c_{\mathrm{max}}\leq15$.
    The NLCE results agree with the DQT and the TCL data at times 
    $t\,\thubbard \lesssim 5$ for both $\varepsilon=0.2$ and $\varepsilon=0.6$.
    Beyond times $t\,\thubbard \sim 5$, the NLCE does not add much to the 
    information on the thermodynamic limit for $\varepsilon=0.2$.
    However, for $\varepsilon=0.6$, the NLCE curves are converged just long enough 
    to indicate that the TCL{\tiny$\infty$} prediction is most suitable to describe 
    $\langle\mathcal{J}(t)\mathcal{J}\rangle$ in the thermodynamic limit.

    \subsubsection{NLCE in a nutshell}
    In the framework of NLCE, the per-site value of the current autocorrelation 
    function on an infinite chain can be expanded in terms of its respective 
    weights $W_c$ on all linked (sub-)clusters (i.e., open-boundary chains of 
    different lengths $c$),
    \begin{align}\label{eq:NLCE}
      \langle\mathcal{J}(t)\mathcal{J}\rangle/L=\sum_cW_c(t)\ .
    \end{align}
    For numerical calculations, the sum in Eq.~\eqref{eq:NLCE} naturally has to 
    be truncated to the maximum accessible cluster size $c_{\mathrm{max}}$. This, 
    together with the inclusion-exclusion principle for the calculation of each weight, 
    $W_c(t)=\langle\mathcal{J}(t)\mathcal{J}\rangle^{(c)}-\sum_{s\subset c}W_s(t)$, 
    results in the very simple expression approximating Eq.~\eqref{eq:NLCE},
    \begin{align}\label{eq:trunc-NLCE}
      \sum_c^{c_{\mathrm{max}}}W_c(t)=
      \langle\mathcal{J}(t)\mathcal{J}\rangle^{(c_{\mathrm{max}})}
      -\langle\mathcal{J}(t)\mathcal{J}\rangle^{(c_{\mathrm{max}}-1)}\ ,
    \end{align}
    which is reliable 
    up to a certain maximum time, increasing with the maximum cluster size 
    $c_{\mathrm{max}}$.
    The $\langle\mathcal{J}(t)\mathcal{J}\rangle^{(c)}$ (evaluated 
    on open-boundary chains of length $c$) are again obtained with DQT and additionally 
    averaged over multiple random states in order to counteract the sensitivity of the 
    difference in Eq.~\eqref{eq:trunc-NLCE} to small statistical errors.
    
    \section{Relation to correlation functions}\label{app:proof_correlationfunc}
    
    To see that $\langle {\cal J} {\cal P} \rho_\mathrm{I}(t) \rangle \propto
      \langle {\cal J}(t) {\cal J}_\mathrm{I}(t) \rangle$, we first insert the  
    definition of the projection superoperator $\cal P$ in $\langle {\cal J}
      {\cal P} \rho_\mathrm{I}(t) \rangle$, which yields
    \begin{equation}
      \langle {\cal J} {\cal P} \rho_\mathrm{I}(t) \rangle = \langle {\cal J}
      \Big( \frac{1}{D} + \frac{\langle {\cal J} \rho_\mathrm{I}(t) \rangle}{\langle 
        {\cal J}^2 \rangle} \, {\cal J} \Big) \rangle \, .
    \end{equation}
    Since $\langle {\cal J} \rangle = 0$, performing the outer angles leads to
    \begin{equation}
      \langle {\cal J} {\cal P} \rho_\mathrm{I}(t) \rangle = \langle {\cal J}
      \rho_\mathrm{I}(t) \rangle \, .
    \end{equation}
    We then insert the initial condition $\rho(0) \propto 1 + b \cal {J}$ and use 
    the time dependence of a density matrix in the interaction picture, 
    $\rho_\mathrm{I}(t) =  e^{i {\cal H}_0 t} \, e^{-i {\cal H} t} \,
      \rho(0) \, e^{i {\cal H} t} \, e^{-i {\cal H}_0 t}$, to obtain
    \begin{equation}
      \langle {\cal J} {\cal P} \rho_\mathrm{I}(t) \rangle \propto \langle 
      {\cal J} e^{i {\cal H}_0 t} \, e^{-i {\cal H} t} \, (1 + b {\cal J}) 
      \, e^{i {\cal H} t} \, e^{-i {\cal H}_0 t} \rangle \, .
    \end{equation}
    Using $\langle {\cal J} \rangle = 0$ again, we thus get
    \begin{equation}
      \langle {\cal J} {\cal P} \rho_\mathrm{I}(t) \rangle \propto \langle 
      {\cal J} e^{i {\cal H}_0 t} \, e^{-i {\cal H} t} \, {\cal J}
      \, e^{i {\cal H} t} \, e^{-i {\cal H}_0 t} \rangle \, ,
    \end{equation}
    which, after a cyclic permutation, reads
    \begin{equation}
      \langle {\cal J} {\cal P} \rho_\mathrm{I}(t) \rangle \propto \langle
      e^{-i {\cal H} t} \, {\cal J} e^{i {\cal H} t} \, e^{-i 
      {\cal H}_0 t} \, {\cal J} \, e^{i {\cal H}_0 t}  \rangle \, .
    \end{equation}
    Denoting by ${\cal J}(t) = e^{i {\cal H} t} \, {\cal J} \, e^{-i {\cal H}
      t}$ and ${\cal J}_\mathrm{I}(t) = e^{i {\cal H}_0 t} \, {\cal J} \, e^{-i
      {\cal H}_0 t}$ the time evolution of an operator in the Heisenberg and 
    interaction picture, respectively, we can write
    \begin{equation}
      \langle {\cal J} {\cal P} \rho_\mathrm{I}(t) \rangle \propto \langle
      {\cal J}(-t) {\cal J}_\mathrm{I}(-t) \rangle \, .
    \end{equation}
    Due to ${\cal J}^\dagger = {\cal J}$, we can replace $t \to -t$ and end up with
    \begin{equation}
      \langle {\cal J} {\cal P} \rho_\mathrm{I}(t) \rangle \propto \langle
      {\cal J}(t) {\cal J}_\mathrm{I}(t) \rangle \, .
    \end{equation}

    \section{Conductivity}\label{app:conductivity}
    
    \begin{figure}[b]
      \centering
      \includegraphics[width=.5\textwidth]{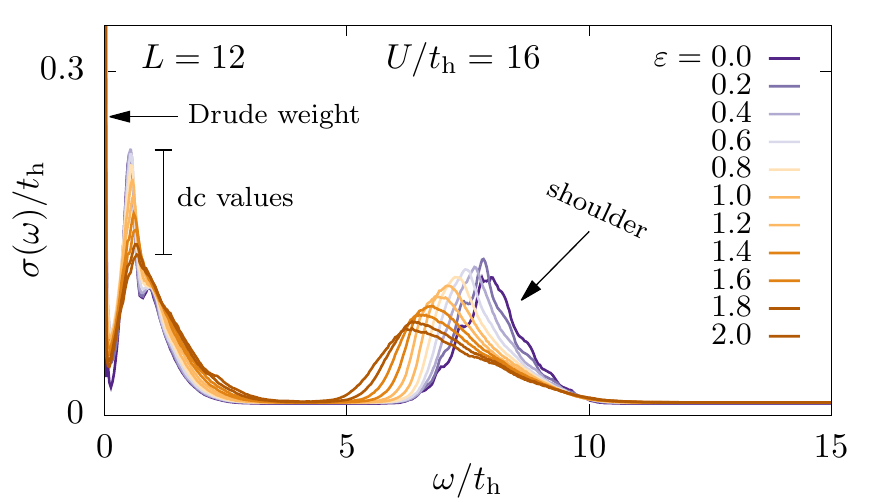}
      \caption{Frequency dependence of the conductivity
        $\sigma(\omega)$, as obtained by the Fourier transform \eqref{supp-eq:FT}
        of the current autocorrelation functions shown in Fig.~\ref{fig:3}(a) of the main text.}
      \label{app-fig:5}
    \end{figure}
    
    In Fig.~\ref{app-fig:5}, we show the frequency-dependent conductivity 
    in the strongly interacting system ${\cal H}(U/t_\mathrm{h} = 16, U' = 0)$, 
    perturbed by interactions $U'/\thubbard=\varepsilon$. This conductivity is
    obtained by the Fourier transform of the current autocorrelation functions 
    depicted in Fig.~\ref{fig:3}(a) of the main text,
    \begin{align}\label{supp-eq:FT}
      \sigma(\omega)= \int^{t_{\mathrm{max}}}_{-t_{\mathrm{max}}}\mathrm{d}t\ 
      \mathrm{e}^{-i\omega t} \, \frac{\langle\mathcal{J}(t)\mathcal{J}\rangle}{L}\ ,
    \end{align}
    with a cutoff time $t_{\mathrm{max}}\,\thubbard=100$. The overall shape of the 
    conductivity is incompatible with a simple Lorentzian form, while 
    the freestanding shoulder shifts from higher to lower $\omega$ as the 
    perturbation strength increases, attesting to the shift in the frequencies 
    observed in the oscillatory behavior of the corresponding current 
    autocorrelation functions. In addition, the spectral weight at small $\omega$
    provides a rough estimate for the value of the dc conductivity $\sigma_{\mathrm
        dc}$, which clearly does not scale as $\sigma_\text{dc} \propto
      1/\varepsilon^2$. For the detailed extraction of $\sigma_\text{dc}$, see Ref.\ 
    \cite{Bertini2020}.
    
  \end{appendix}


\bibliographystyle{apsrev4-1}

\bibliography{nontrivial}

\end{document}